# An Achievable Rate Region for Cognitive Radio Channel With Common Message

M. Monemizadeh, G. Abed Hodtani, and H. Fehri
Department of Electrical Engineering
Ferdowsi University of Mashhad
Mashhad, Iran
mostafamonemizadeh@gmail.com, ghodtani@gmail.com, hamed_fehri@yahoo.com

*Abstract*—The cognitive radio channel with common message (CRCC) is considered. In this channel, similar to the cognitive radio channel (CRC), we have a cognitive user which has full non-causal knowledge of the primary message, and like the interference channel with common message (ICC), the information sources at the two transmitters are statistically dependent and the senders need to transmit not only the private message but also certain common message to their corresponding receivers. By using a specific combination of superposition coding, binning scheme and simultaneous decoding, we propose a unified achievable rate region for the CRCC which subsumes the several existing results for the CRC, ICC, interference channel without common message (IC), strong interference channel and compound multiple access channel with common information (SICC and CMACC).

## I. INTRODUCTION

Two-sender, two-receiver channel models allow various forms of transmitter cooperation. When senders are unaware of each other's messages, we have the interference channel (IC) [1]–[5]. In the IC, the level of cooperation between transmitters and the resulting performance improvement will depend on the amount of information that encoders share [6].

Channel models with cooperating nodes are of interest also for networks with cognitive users. The cognitive radio channel (CRC) is a well studied channel model inspired by the potential abilities of cognitive radio technology and its impact on spectral efficiency in wireless networks. The CRC consists of a classical two-user interference channel with two independent messages, in which the message of the "primary" user is non-causally provided to the transmitter of the other "secondary" or "cognitive" user, while the primary transmitter has the knowledge of its own message only. From an information-theoretic perspective, the CRC has been investigated in [4], [6]–[11].

The assumption of the statistical independence of the source messages becomes invalid in two-sender, two-receiver channel where the senders need transmit not only the private message but also certain common message to their corresponding receivers. Such a scenario can be modeled as the IC with common message (ICC) ([12] and [13]) or in more general case it can be modeled as the CRC with common message (CRCC). In this paper, we consider a more general case than the ICC and the CRC which can be modeled as the CRC with common message (CRCC). By using a specific combination of superposition coding, binning scheme and simultaneous decoding, we propose an achievable rate region for the CRCC which subsumes the some existing results for the CRC, ICC, IC, the strong interference channel and compound multiple access channel with common message (SICC and CMACC). The rest of the paper is organized as follows. In Section II, we define CRCC and the modified CRCC. In Section III, we present the achievable rate region for the general discrete memoryless CRCC. In Section IV, we discuss some ramifications of the main results. Finally, a conclusion is prepared in Section V.

## II. DEFINITIONS

We denote random variables by $X_1, X_2, Y_1, \cdots$ with values $x_1, x_2, y_1, \cdots$ in finite sets $\mathcal{X}_1, \mathcal{X}_2, \mathcal{Y}_1, \cdots$ respectively; n-tuple vectors of $X_1, X_2, Y_1, \cdots$ are denoted with $\mathbf{x_1}, \mathbf{x_2}, \mathbf{y_1}, \cdots$.

*Definition 1 (Two-User CRCC):* A two-user discrete memoryless CRCC is usually defined by a quintuple $(\mathcal{X}_1, \mathcal{X}_2, \mathcal{P}, \mathcal{Y}_1, \mathcal{Y}_2)$ where $\mathcal{X}_t$ and $\mathcal{Y}_t$, $t = 1,2$, denote the finite channel input and output alphabets respectively, and $\mathcal{P}$ denotes the collection of the conditional probabilities $p(y_1, y_2 | x_1, x_2)$ on $(y_1, y_2) \in \mathcal{Y}_1 \times \mathcal{Y}_2$ given $(x_1, x_2) \in \mathcal{X}_1 \times \mathcal{X}_2$. As shown in Fig. 1, sender $t$, $t = 1,2$, sends a private message $m_t \in \mathcal{M}_t = \{1, \cdots, M_t\}$ together with a common message $m_0 \in \mathcal{M}_0 = \{1, \cdots, M_0\}$ to its pairing receiver. Note that the second sender is non-causally aware of the message to be sent by the first sender.

Let $\mathcal{C}$ denote the discrete memoryless CRCC defined above. An $(n, M_0 = \lfloor 2^{nR_0} \rfloor, M_1 = \lfloor 2^{nR_1} \rfloor, M_2 = \lfloor 2^{nR_2} \rfloor, \varepsilon)$

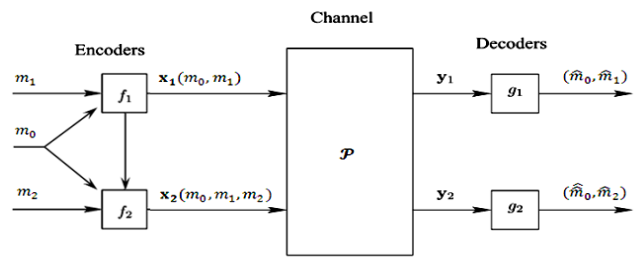

Figure 1.  Cognitive radio channel with common message.

code exists for the channel $\mathcal{C}$ if and only if there exist two encoding functions

$$f_1: \mathcal{M}_0 \times \mathcal{M}_1 \to \mathcal{X}_1^n, \qquad f_2: \mathcal{M}_0 \times \mathcal{M}_1 \times \mathcal{M}_2 \to \mathcal{X}_2^n$$

and two decoding functions

$$g_1: \mathcal{Y}_1^n \to \mathcal{M}_0 \times \mathcal{M}_1, \qquad g_2: \mathcal{Y}_2^n \to \mathcal{M}_0 \times \mathcal{M}_2$$

such that $max(P_{e1}^n, P_{e2}^n) \leq \varepsilon$, where $P_{e1}^n$ and $P_{e2}^n$ are the average error probabilities of decoding error. A nonnegative rate triple $(R_0, R_1, R_2)$ is achievable for the channel $\mathcal{C}$ if for any given $0 < \varepsilon < 1$, and for any sufficiently large $n$, there exists a code $(n, \lfloor 2^{nR_0} \rfloor, \lfloor 2^{nR_1} \rfloor, \lfloor 2^{nR_2} \rfloor, \varepsilon)$. The capacity region for the channel $\mathcal{C}$ is defined as the closure of the set of all the achievable rate triples.

*Definition 2 (Modified CRCC):* A modified CRCC (Fig.2), has five streams of messages instead of three in the associated CRCC. The five streams of messages $n_0, n_1, l_1, n_2$, and $l_2$ are assumed to be independently and uniformly generated over the finite sets $\mathcal{N}_0 = \{1, \cdots, N_0\}$, $\mathcal{N}_1 = \{1, \cdots, N_1\}$, $\mathcal{L}_1 = \{1, \cdots, L_1\}$, $\mathcal{N}_2 = \{1, \cdots, N_2\}$, and $\mathcal{L}_2 = \{1, \cdots, L_2\}$, respectively. Denote the modified CRCC by $\mathcal{C}_m$.

An $(n, N_0 = \lfloor 2^{nT_0} \rfloor, N_1 = \lfloor 2^{nT_1} \rfloor, L_1 = \lfloor 2^{nS_1} \rfloor, N_2 = \lfloor 2^{nT_2} \rfloor, L_2 = \lfloor 2^{nS_2} \rfloor, \varepsilon)$ code exists for the channel $\mathcal{C}_m$ if and only if there exist two encoding functions

$$f_1: \mathcal{N}_0 \times \mathcal{N}_1 \times \mathcal{L}_1 \to \mathcal{X}_1^n,$$
$$f_2: \mathcal{N}_0 \times \mathcal{N}_1 \times \mathcal{L}_1 \times \mathcal{N}_2 \times \mathcal{L}_2 \to \mathcal{X}_2^n$$

and two decoding functions

$$g_1: \mathcal{Y}_1^n \to \mathcal{N}_0 \times \mathcal{N}_1 \times \mathcal{L}_1 \times \mathcal{N}_2,$$
$$g_2: \mathcal{Y}_2^n \to \mathcal{N}_0 \times \mathcal{N}_2 \times \mathcal{L}_2 \times \mathcal{N}_1$$

such that $max(P_{e1}^n, P_{e2}^n) \leq \varepsilon$, where $P_{e1}^n$ and $P_{e2}^n$ are the average error probabilities of decoding error. A nonnegative rate quintuple $(T_0, T_1, S_1, T_2, S_2)$ is achievable for the channel $\mathcal{C}_m$ if for any given $0 < \varepsilon < 1$, and for any sufficiently large $n$, there exists a code $(n, \lfloor 2^{nT_0} \rfloor, \lfloor 2^{nT_1} \rfloor, \lfloor 2^{nS_1} \rfloor, \lfloor 2^{nT_2} \rfloor, \lfloor 2^{nS_2} \rfloor, \varepsilon)$ for the channel $\mathcal{C}_m$.

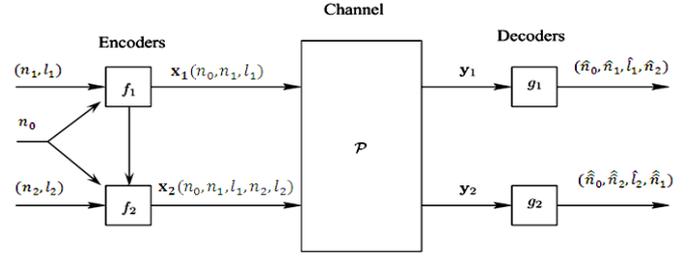

Figure 2. Modified cognitive radio channel with common message.

### III. MAIN RESULTS

#### A. An Achievable Rate Region for the CRCC

Let us consider auxiliary random variables $W_0, W_1, U_1, W_2$, and $U_2$ defined on arbitrary finite sets $\mathcal{W}_0, \mathcal{W}_1, \mathcal{U}_1, \mathcal{W}_2$, and $\mathcal{U}_2$, respectively. For the modified CRCC (Fig. 2), let $Z = (W_0 W_1 U_1 W_2 U_2 X_1 X_2 Y_1 Y_2)$ and let $\mathcal{P}_{CRCC}$ be the set of all distributions of the form (1):

$$p(w_0 w_1 u_1 w_2 u_2 x_1 x_2 y_1 y_2) = p(w_0) p(w_1 u_1 | w_0)$$
$$\cdot p(w_2 u_2 | w_0 w_1 u_1) p(x_1 x_2 | w_0 w_1 u_1 w_2 u_2) p(y_1 y_2 | x_1 x_2) \quad (1)$$

By considering the general distribution (1) for the CRCC which allows the auxiliary variables to be correlated in the form (2):

$$p(w_0 w_1 u_1 w_2 u_2) = p(w_0) p(w_1 | w_0) p(u_1 | w_0 w_1)$$
$$\cdot p(w_2 | w_0 w_1 u_1) p(u_2 | w_0 w_2 w_1 u_1) \quad (2)$$

and using a specific combination of superposition coding, binning scheme and simultaneous decoding, we obtain a unified achievable rate region for the general CRCC as follows.

*Theorem 1:* For the modified CRCC (Fig. 2), let $Z = (W_0 U_1 W_1 U_2 W_2 X_1 X_2 Y_1 Y_2)$ and let $\mathcal{P}_{CRCC}$ be the set of all distributions of the form (2). For any $Z \in \mathcal{P}_{CRCC}$, let $S_{CRCC}(Z)$ be the set of all non-negative quintuples $(T_0, T_1, S_1, T_2, S_2)$ satisfying (3-1)-(3-16) (at the bottom of this page), then any element of the closure of $\bigcup_{Z \in \mathcal{P}_{CRCC}} S_{CRCC}(Z)$ is achievable.

*Proof:* Refer to Appendix A.

It is easy to prove that the rate region $S_{CRCC}$ is convex.

$$S_1 \leq I(Y_1; U_1|W_0 W_1 W_2) + I(W_2; W_1 U_1|W_0) = A_1 \tag{3-1}$$
$$T_1 \leq I(Y_1; W_1|W_0 U_1 W_2) + I(U_1; W_1|W_0) + I(W_2; W_1 U_1|W_0) = B_1 \tag{3-2}$$
$$T_2 \leq I(Y_1; W_2|W_0 W_1 U_1) + I(U_1; W_1|W_0) = C_1 \tag{3-3}$$
$$S_1 + T_1 \leq I(Y_1; U_1 W_1|W_0 W_2) + I(W_2; W_1 U_1|W_0) = D_1 \tag{3-4}$$
$$S_1 + T_2 \leq I(Y_1; W_2 U_1|W_0 W_1) = E_1 \tag{3-5}$$
$$T_1 + T_2 \leq I(Y_1; W_1 W_2|W_0 U_1) + I(U_1; W_1|W_0) = F_1 \tag{3-6}$$
$$S_1 + T_1 + T_2 \leq I(Y_1; W_1 W_2 U_1|W_0) = G_1 \tag{3-7}$$
$$S_1 + T_0 + T_1 + T_2 \leq I(Y_1; W_0 W_1 W_2 U_1) = H_1 \tag{3-8}$$
$$S_2 \leq I(Y_2; U_2|W_0 W_1 W_2) + I(U_2; W_2|W_0) + I(W_1; W_2 U_2|W_0) - I(U_2; U_1 W_1 W_2|W_0) = A_2 \tag{3-9}$$
$$T_2 \leq I(Y_2; W_2|W_0 W_1 U_2) + I(U_2; W_2|W_0) + I(W_1; W_2 U_2|W_0) - I(W_2; W_1 U_1|W_0) = B_2 \tag{3-10}$$
$$T_1 \leq I(Y_2; W_1|W_0 W_2 U_2) + I(U_2; W_2|W_0) + I(W_1; W_2 U_2|W_0) = C_2 \tag{3-11}$$
$$S_2 + T_2 \leq I(Y_2; U_2 W_2|W_0 W_1) + I(U_2; W_2|W_0) + I(W_1; W_2 U_2|W_0) - I(W_2; W_1 U_1|W_0) - I(U_2; U_1 W_1 W_2|W_0) = D_2 \tag{3-12}$$
$$S_2 + T_1 \leq I(Y_2; U_2 W_1|W_0 W_2) + I(U_2; W_2|W_0) + I(W_1; W_2 U_2|W_0) - I(U_2; U_1 W_1 W_2|W_0) = E_2 \tag{3-13}$$
$$T_2 + T_1 \leq I(Y_2; W_1 W_2|U_2 W_0) + I(U_2; W_2|W_0) + I(W_1; W_2 U_2|W_0) - I(W_2; W_1 U_1|W_0) = F_2 \tag{3-14}$$
$$S_2 + T_2 + T_1 \leq I(Y_2; W_1 W_2 U_2|W_0) + I(U_2; W_2|W_0) + I(W_1; W_2 U_2|W_0) - I(W_2; W_1 U_1|W_0) - I(U_2; U_1 W_1 W_2|W_0) = G_2 \tag{3-15}$$
$$S_2 + T_0 + T_2 + T_1 \leq I(Y_2; W_0 W_1 W_2 U_2) + I(U_2; W_2|W_0) + I(W_1; W_2 U_2|W_0) - I(W_2; W_1 U_1|W_0) - I(U_2; U_1 W_1 W_2|W_0) = H_2 \tag{3-16}$$

## B. Explicit Description of the Achievable Rate Region

Now, we describe the above region in theorem 1 in terms of the rate triples $(R_0 = T_0, R_1 = S_1 + T_1, R_2 = S_2 + T_2)$ by the Fourier-Motzkin elimination technique.

*Theorem 2:* The $\mathcal{S}_{CRCC}$ region in theorem 1 can be described as $\mathcal{R}_{CRCC}$ being the set of $(R_0, R_1, R_2)$ satisfying:

$$R_1 \leq D_1 \tag{4-1}$$
$$R_1 \leq G_1 \tag{4-2}$$
$$R_1 \leq A_1 + C_2 \tag{4-3}$$
$$R_1 \leq A_1 + E_2 \tag{4-4}$$
$$R_1 \leq B_1 + E_1 \tag{4-5}$$
$$R_1 \leq A_1 + F_2 \tag{4-6}$$
$$R_1 \leq E_1 + C_2 \tag{4-7}$$
$$R_1 \leq E_1 + F_2 \tag{4-8}$$
$$R_2 \leq D_2 \tag{4-9}$$
$$R_2 \leq A_2 + C_1 \tag{4-10}$$
$$R_2 \leq A_2 + E_1 \tag{4-11}$$
$$R_0 + R_1 \leq H_1 \tag{4-12}$$
$$R_0 + R_2 \leq H_2 \tag{4-13}$$
$$R_1 + R_2 \leq A_1 + G_2 \tag{4-14}$$
$$R_1 + R_2 \leq E_1 + E_2 \tag{4-15}$$
$$R_1 + R_2 \leq A_2 + G_1 \tag{4-16}$$
$$R_1 + R_2 \leq E_1 + G_2 \tag{4-17}$$
$$R_1 + R_2 \leq A_2 + B_1 + E_1 \tag{4-18}$$
$$2R_1 + R_2 \leq 2A_1 + E_2 + F_2 \tag{4-19}$$
$$2R_1 + R_2 \leq A_1 + E_2 + G_1 \tag{4-20}$$
$$R_1 + 2R_2 \leq 2A_2 + E_1 + F_1 \tag{4-21}$$
$$R_1 + 2R_2 \leq A_2 + E_1 + G_2 \tag{4-22}$$
$$R_0 + R_1 + R_2 \leq A_1 + H_2 \tag{4-23}$$
$$R_0 + R_1 + R_2 \leq A_2 + H_1 \tag{4-24}$$
$$R_0 + R_1 + R_2 \leq E_1 + H_2 \tag{4-25}$$
$$R_0 + 2R_1 + R_2 \leq A_1 + E_2 + H_1 \tag{4-26}$$
$$R_0 + R_1 + 2R_2 \leq A_2 + E_1 + H_2 \tag{4-27}$$

where the bound constants $A_i, B_i, C_i, D_i, E_i, F_i, G_i, H_i, i = 1,2$ are the same as in theorem 1.

*Proof:* Refer to Appendix B.

## IV. RAMIFICATIONS OF MAIN RESULTS

In this section, we discuss some special cases of the proposed achievable rate region to demonstrate the breadth of our main results.

### A. Interference Channel Without Common Message (IC)

We now consider the general IC as a special case of the CRCC. Let $Q \in \mathcal{Q}$ denote the time-sharing random variable whose n-sequences $\mathbf{q} = (q_1, q_2, \cdots, q_n)$ are generated independently of the messages. The n-sequences $\mathbf{q}$ are given to both senders and receivers. Since no common message is involved for the IC, we can choose $W_0 = Q$ and $T_0 = 0$.

*Remark 1:* By setting $W_0 = Q, T_0 = 0$, and considering the below distribution instead of (2),
$p(qw_1u_1w_2u_2) = p(q)p(w_1|q)p(u_1|qw_1)p(w_2|q)p(u_2|qw_2)$
the rate region in theorem 1 is reduced to the Hodtani region for the IC (theorem 2, [4] and theorem 7, [5]) in terms of the quadruple $(S_1, T_1, S_2, T_2)$, and the rate region in theorem 2 is reduced to the Hodtani region for the IC (theorem 8, [5]) in terms of the $(R_1, R_2)$.

*Remark 2:* By setting $W_0 = Q, T_0 = 0$, and considering the below distribution instead of (2),
$p(qw_1u_1w_2u_2) = p(q)p(w_1|q)p(u_1|q)p(w_2|q)p(u_2|q)$
the rate region in theorem 1 is reduced to the HK region (theorem 3.1, [3]) in terms of the quadruple $(S_1, T_1, S_2, T_2)$, and the HK region for the IC in terms of the rate pair $(R_1, R_2)$ (theorem 2, [5]) is readily derived from theorem 2.

### B. Interference Channel With Common Message (ICC)

*Remark 3:* By considering the below distribution instead of (2),
$p(w_0w_1u_1w_2u_2) = p(w_0)p(w_1|w_0)$
$\quad\quad . p(u_1|w_0w_1)p(w_2|w_0)p(u_2|w_0w_2)$
the rate region in theorem 1 is reduced to the rate region for the ICC.

### C. Cognitive Radio Channel Without Common Message

*Remark 4:* By setting $W_0 = Q, T_0 = 0$, and considering the below distribution instead of (2),
$p(qw_1u_1w_2u_2) =$
$\quad p(q)p(w_1|q)p(u_1|qw_1)p(w_2|qu_1w_1)p(u_2|qu_1w_1w_2)$
the rate region in theorem 1 is reduced to the Hodtani region for CRC (theorem 3, [4] and theorem 3, [11]) in terms of the quadruple $(S_1, T_1, S_2, T_2)$, and the rate region in theorem 2 is reduced to the Hodtani region for the CRC (theorem 4, [11]) in terms of the $(R_1, R_2)$.

### D. Compound Multiple Access Channel With Common Message (CMACC)

Let $\mathcal{P}_{CMACC}$ be the set of all distributions of the form (5):
$$p(w_0w_1w_2x_1x_2y_1y_2) = p(w_0)p(w_1|w_0)\,p(w_2|w_0)$$
$$\quad . p(x_1|w_0w_1)p(x_2|w_0w_2)p(y_1y_2|x_1x_2) \tag{5}$$

*Remark 5:* For the CMACC, let $Z_M = (W_0W_1W_2X_1X_2Y_1Y_2)$ and let $\mathcal{P}_{CMACC}$ be the set of all distributions of the form (5). For any $Z_M \in \mathcal{P}_{CMACC}$, let $\mathcal{S}_{CMACC}(Z_M)$ be the set of all nonnegative rate triples $(R_0, R_1, R_2)$ such that

$$R_1 \leq \min\{I(Y_1; W_1|W_0W_2), I(Y_2; W_1|W_0W_2)\} \tag{6-1}$$
$$R_2 \leq \min\{I(Y_1; W_2|W_0W_1), I(Y_2; W_2|W_0W_1)\} \tag{6-2}$$
$$R_1 + R_2 \leq \min\{I(Y_1; W_1W_2|W_0), I(Y_2; W_1W_2|W_0)\} \tag{6-3}$$
$$R_0 + R_1 + R_2 \leq \min\{I(Y_1; W_0W_1W_2), I(Y_2; W_0W_1W_2)\} \tag{6-4}$$

then any element of the closure of $\bigcup_{Z_M \in \mathcal{P}_{CMACC}} \mathcal{S}_{CMACC}(Z_M)$ is achievable (by letting $X_i = W_i, i = 1,2$; [16, Achievability of theorem 1]).

*Remark 6:* By setting $U_i = \emptyset$, $S_i = 0$, $i = 1,2$; $R_0 = T_0, R_1 = T_1, R_2 = T_2$ in (3-1)–(3-16), considering the (5) instead of (2), and removing redundant ones from the resulting inequalities, we can easily obtain (6-1)–(6-4).

### E. Strong Interference Channel With Common Message

Let $\mathcal{P}_{SICC}$ be the set of all distributions of the form (5). We can consider an ICC as a SICC if

$$I(W_1; Y_1|W_2W_0) \leq I(W_1; Y_2|W_2W_0) \tag{7-1}$$
$$I(W_2; Y_2|W_1W_0) \leq I(W_2; Y_1|W_1W_0). \tag{7-2}$$

*Remark 7:* For the SICC, let $Z_s = (W_0W_1W_2X_1X_2Y_1Y_2)$ and let $\mathcal{P}_{SICC}$ be the set of all distributions of the form (5).

For any $Z_s \in \mathcal{P}_{SICC}$, let $\mathcal{S}_{SICC}(Z_s)$ be the set of all nonnegative rate triples $(R_0, R_1, R_2)$ such that

$$R_1 \leq I(Y_1; W_1|W_0 W_2) \quad (8\text{-}1)$$
$$R_2 \leq I(Y_2; W_2|W_0 W_1) \quad (8\text{-}2)$$
$$R_1 + R_2 \leq \min\{I(Y_1; W_1 W_2|W_0), I(Y_2; W_1 W_2|W_0)\} \quad (8\text{-}3)$$
$$R_0 + R_1 + R_2 \leq \min\{I(Y_1; W_0 W_1 W_2), I(Y_2; W_0 W_1 W_2)\} \quad (8\text{-}4)$$

then any element of the closure of $\bigcup_{Z_s \in \mathcal{P}_{SICC}} \mathcal{S}_{SICC}(Z_s)$ is achievable (by letting $X_i = W_i$, $i = 1,2$; [16, Achievability of theorem 3]).

*Remark 8:* By setting $U_i = \emptyset$, $S_i = 0$, $i = 1,2$; $R_0 = T_0, R_1 = T_1, R_2 = T_2$ in (3-1)–(3-16), considering the (5) instead of (2), and removing redundant ones from the resulting inequalities due to the channel assumptions of the SICC (i.e. (7-1), (7-2)), we can easily obtain (8-1)–(8-4).

## V. CONCLUSION

By using a specific combination of superposition coding, Gel'fand-Pinsker binning scheme and the HK jointly decoding, we obtained a unified achievable rate region for cognitive radio channel with common message. We have shown that the derived rate region reduces to some existing results developed for the IC, ICC, CRC, SICC and CMACC.

## APPENDIX A
## PROOF OF THEOREM 1

We prove the achievability of any element of $\mathcal{S}_{CRCC}(Z)$ for each $Z \in \mathcal{P}_{CRCC}$. Fix $Z = (W_0 W_1 U_1 W_2 U_2 X_1 X_2 Y_1 Y_2)$.

**Codebook generation:** Consider $n > 0$, some distribution of the form (2) and

$$p(u_1|w_0) = \sum_{w_1} p(w_1|w_0) p(u_1|w_0 w_1) \quad (A\text{-}a)$$

$$p(w_2|w_0) = \sum_{w_1 u_1} p(w_1|w_0) p(u_1|w_0 w_1) p(w_2|w_0 w_1 u_1) \quad (A\text{-}b)$$

$$p(u_2|w_0) = \sum_{w_1 u_1 w_2} \{p(w_1|w_0) p(u_1|w_0 w_1) \cdot p(w_2|w_0 w_1 u_1) p(u_2|w_0 w_1 u_1 w_2)\} \quad (A\text{-}c)$$

First by using superposition coding scheme, we generate $(\boldsymbol{w_1}, \boldsymbol{u_1})$ and $(\boldsymbol{w_2}, \boldsymbol{u_2})$ independently around of $\boldsymbol{w_0}$ (cloud center), and then in accordance with (A-a), (A-b) and (A-c), we can use random binning scheme to generate the sequences of $\boldsymbol{u_1}, \boldsymbol{w_2}$, and $\boldsymbol{u_2}$ independently of $\boldsymbol{w_1}$, $(\boldsymbol{w_1}, \boldsymbol{u_1})$, and $(\boldsymbol{w_1}, \boldsymbol{u_1}, \boldsymbol{w_2})$, respectively. So,

1. generate $\lfloor 2^{nT_0} \rfloor$ independent codewords $\boldsymbol{w_0}(j)$, $j \in \{1, \cdots, \lfloor 2^{nT_0} \rfloor\}$ according to $\prod_{i=1}^{n} p(w_{0i})$.
2. For each codeword $\boldsymbol{w_0}(j)$, at encoder 1:
   2.1 generate $\lfloor 2^{nT_1} \rfloor$ independent codewords $\boldsymbol{w_1}(j,k)$, $k \in \{1, \cdots, \lfloor 2^{nT_1} \rfloor\}$, according to $\prod_{i=1}^{n} p(w_{1i}|w_{0i})$.
   2.2 generate $\lfloor 2^{nz_1} \rfloor$ independent codewords $\boldsymbol{u_1}(j,l)$, $l \in \{1, \cdots, \lfloor 2^{nz_1} \rfloor\}$, according to $\prod_{i=1}^{n} p(u_{1i}|w_{0i})$ and throw them randomly into $\lfloor 2^{nS_1} \rfloor$ bins such that the sequence $\boldsymbol{u_1}(j,l)$ in bin $b_1$ is denoted as $\boldsymbol{u_1}(j,b_1,l)$, $b_1 \in \{1, \cdots, \lfloor 2^{nS_1} \rfloor\}$.
3. For each codeword $\boldsymbol{w_0}(j)$, at encoder 2:
   3.1 generate $\lfloor 2^{nt_2} \rfloor$ independent codewords $\boldsymbol{w_2}(j,m)$, $m \in \{1, \cdots, \lfloor 2^{nt_2} \rfloor\}$, according to $\prod_{i=1}^{n} p(w_{2i}|w_{0i})$ and throw them randomly into $\lfloor 2^{nT_2} \rfloor$ bins such that the sequence $\boldsymbol{w_2}(j,m)$ in bin $b_2$ is denoted as $\boldsymbol{w_2}(j,b_2,m)$, $b_2 \in \{1, \cdots, \lfloor 2^{nT_2} \rfloor\}$.
   3.2 generate $\lfloor 2^{nz_2} \rfloor$ independent codewords $\boldsymbol{u_2}(j,h)$, $h \in \{1, \cdots, \lfloor 2^{nz_2} \rfloor\}$, according to $\prod_{i=1}^{n} p(u_{2i}|w_{0i})$ and throw them randomly into $\lfloor 2^{nS_2} \rfloor$ bins such that the sequence $\boldsymbol{u_2}(j,h)$ in bin $b_3$ is denoted as $\boldsymbol{u_2}(j,b_3,h)$, $b_3 \in \{1, \cdots, \lfloor 2^{nS_2} \rfloor\}$.

**Encoding & transmission:** The messages of $(w_0, w_1, u_1)$ of $(\mathcal{W}_0^n, \mathcal{W}_1^n, \mathcal{U}_1^n)$ and $(w_0, w_2, u_2)$ of $(\mathcal{W}_0^n, \mathcal{W}_2^n, \mathcal{U}_2^n)$ sets are mapped into $\boldsymbol{x_1}$ and $\boldsymbol{x_2}$, respectively through deterministic encoding functions $f_1$ and $f_2$ (as in [3]). The sender $TX_1$ to send $(j,k,b_1)$, looks for $\boldsymbol{w_0}(j), \boldsymbol{w_1}(j,k)$ and finds $\boldsymbol{w_1}(j,k)$; then finds a sequence $\boldsymbol{u_1}(j,l)$ in bin $b_1$ such that $(\boldsymbol{w_0}(j), \boldsymbol{w_1}(j,k), \boldsymbol{u_1}(j,b_1,l)) \in A_\varepsilon^n$; finally sender $TX_1$ generates $\boldsymbol{x_1}$ i.i.d. according to $x_{1i} = f_1(w_{0i}(j), w_{1i}(j,k), u_{1i}(j,b_1,l))$, $i = 1, \ldots, n$; and sends it. The sender $TX_2$ to send $(j,b_2,b_3)$, being non-causally aware of $\boldsymbol{w_1}(j,k), \boldsymbol{u_1}(j,b_1,l)$, looks for $\boldsymbol{w_0}(j)$ and finds a sequence $\boldsymbol{w_2}(j,m)$ in bin $b_2$ such that $(\boldsymbol{w_0}(j), \boldsymbol{w_1}(j,k), \boldsymbol{u_1}(j,b_1,l), \boldsymbol{w_2}(j,b_2,m)) \in A_\varepsilon^n$; then finds a sequence $\boldsymbol{u_2}(j,h)$ in bin $b_3$ such that $(\boldsymbol{w_0}(j), \boldsymbol{w_1}(j,k), \boldsymbol{u_1}(j,b_1,l), \boldsymbol{w_2}(j,b_2,m), \boldsymbol{u_2}(j,b_3,h)) \in A_\varepsilon^n$; finally sender $TX_2$ generates $\boldsymbol{x_2}$ i.i.d. according to $x_{2i} = f_2(w_{0i}(j), w_{2i}(j,b_2,m), u_{2i}(j,b_3,h)), i = 1, \ldots, n$ ; and sends it.

**Error probability analysis:** The messages are decoded based on strong joint typicality as in [3]. Assuming all messages to be equiprobable, we consider the situation where $(j = 1, k = 1, b_1 = 1, b_2 = 1, b_3 = 1)$ was sent. Here, we do the error analysis for the second receiver $RX_2$. The receiver $RX_2$ upon receiving $\boldsymbol{y_2}$ decodes $jk(b_2, m)(b_3, h) = 11(1, m)(1, h)$ or $(j = 1, k = 1, b_2 = 1, b_3 = 1)$ simultaneously [3]. We define the error event $E^2_{jk(b_2,m)(b_3,h)}$ and $P^n_{e2}$ as follows.

$E^2_{jk(b_2,m)(b_3,h)} = \{(\boldsymbol{w_0}(j), \boldsymbol{w_1}(j,k), \boldsymbol{w_2}(j,b_2,m), \boldsymbol{u_2}(j,b_3,h), \boldsymbol{y_2}) \in A_\varepsilon^n\}$

$P^n_{e2} = P\{E^{2C}_{11(1,m)(1,h)} \cup E^2_{jk(b_2,m)(b_3,h) \neq 11(1,m)(1,h)}\} \leq$

$P(E^{2C}_{11(1,m)(1,h)}) + \sum_{jk(b_2,m)(b_3,h) \neq 11(1,m)(1,h)} P(E^2_{jk(b_2,m)(b_3,h)})$

$$\leq \varepsilon + \underbrace{\sum_{\substack{j \neq 1, \\ k = b_2 = b_3 = 1}}}_{p_1} \cdots + \underbrace{\sum_{\substack{k \neq 1, \\ j = b_2 = b_3 = 1}}}_{p_2} \cdots + \underbrace{\sum_{\substack{b_2 \neq 1, \\ j = k = b_3 = 1}}}_{p_3} \cdots +$$

$$\underbrace{\sum_{\substack{b_3 \neq 1, \\ j = k = b_2 = 1}}}_{p_4} \cdots + \underbrace{\sum_{\substack{j \neq 1, k \neq 1, \\ b_2 = b_3 = 1}}}_{p_5} \cdots + \underbrace{\sum_{\substack{j \neq 1, b_2 \neq 1, \\ k = b_3 = 1}}}_{p_6} \cdots + \underbrace{\sum_{\substack{j \neq 1, b_3 \neq 1, \\ k = b_2 = 1}}}_{p_7} \cdots +$$

$$\underbrace{\sum_{\substack{k \neq 1, b_2 \neq 1, \\ j = b_3 = 1}}}_{p_8} \cdots + \underbrace{\sum_{\substack{k \neq 1, b_3 \neq 1, \\ j = b_2 = 1}}}_{p_9} \cdots + \underbrace{\sum_{\substack{b_2 \neq 1, b_3 \neq 1, \\ j = k = 1}}}_{p_{10}} \cdots + \underbrace{\sum_{\substack{j \neq 1, k \neq 1, \\ b_2 \neq 1, b_3 = 1}}}_{p_{11}} \cdots +$$

$$\sum_{\substack{j\neq 1, k\neq 1,\\ b_3\neq 1, b_2=1}}^{\boxed{p_{12}}} \cdots + \sum_{\substack{j\neq 1, b_2\neq 1,\\ b_3\neq 1, k=1}}^{\boxed{p_{13}}} \cdots + \sum_{\substack{k\neq 1, b_2\neq 1,\\ b_3\neq 1, j=1}}^{\boxed{p_{14}}} \cdots + \sum_{\substack{j\neq 1, k\neq 1,\\ b_2\neq 1, b_3\neq 1}}^{\boxed{p_{15}}} \cdots$$

Let us choose $\boxed{p_7}$, $\boxed{p_{14}}$ for evaluation. So, according to the codebook generation and the original distribution (2) in theorem 1, we have:

■ $\sum_{j\neq 1, b_3\neq 1, k=b_2=1}^{\boxed{p_7}} \cdots \leq 2^{n(T_0+z_2)} \times$
$P\{(\boldsymbol{w_0}(j), \boldsymbol{w_1}(j,1), \boldsymbol{w_2}(j,1,m), \boldsymbol{u_2}(j,b_3,h), \boldsymbol{y_2}) \in A_\varepsilon^n\}$
$\leq 2^{n(T_0+z_2)} \|A_\varepsilon^n\| \, p(\boldsymbol{w_0})p(\boldsymbol{w_1}|\boldsymbol{w_0})p(\boldsymbol{w_2}|\boldsymbol{w_0})p(\boldsymbol{u_2}|\boldsymbol{w_0})p(\boldsymbol{y_2})$
$\leq 2^{n(T_0+z_2)} \times (2^{nH(W_0 W_1 W_2 U_2 Y_2)}) \times (2^{-nH(W_0)}$
$\quad \cdot 2^{-nH(W_1|W_0)} \cdot 2^{-nH(W_2|W_0)} \cdot 2^{-nH(U_2|W_0)} \cdot 2^{-nH(Y_2)})$
$= 2^{n(T_0+z_2)} \times$
$2^{n(H(W_0)+H(W_1|W_0)+H(W_2|W_0 W_1)+H(U_2|W_0 W_1 W_2)+H(Y_2|W_0 W_1 W_2 U_2))}$
$\times 2^{-n(H(W_0)+H(W_1|W_0)+H(W_2|W_0)+H(U_2|W_0)+H(Y_2))}$
$= 2^{-n(I(Y_2;W_0 W_1 W_2 U_2)+I(U_2;W_2|W_0)+I(W_1;W_2 U_2|W_0)-(T_0+z_2))}$

■ $\sum_{k\neq 1, b_2\neq 1, b_3\neq 1, j=1}^{\boxed{p_{14}}} \cdots \leq 2^{n(T_1+t_2+z_2)} \times$
$P\{(\boldsymbol{w_0}(1), \boldsymbol{w_1}(1,k), \boldsymbol{w_2}(1,b_2,m), \boldsymbol{u_2}(1,b_3,h), \boldsymbol{y_2}) \in A_\varepsilon^n\}$
$\leq 2^{n(T_1+t_2+z_2)} \times \|A_\varepsilon^n\| \times$
$p(\boldsymbol{w_0})p(\boldsymbol{w_1}|\boldsymbol{w_0})p(\boldsymbol{w_2}|\boldsymbol{w_0})p(\boldsymbol{u_2}|\boldsymbol{w_0})p(\boldsymbol{y_2}|\boldsymbol{w_0})$
$\leq 2^{n(T_1+t_2+z_2)} \times (2^{nH(W_0 W_1 W_2 U_2 Y_2)}) \times (2^{-nH(W_0)}$
$\quad \cdot 2^{-nH(W_1|W_0)} \cdot 2^{-nH(W_2|W_0)} \cdot 2^{-nH(U_2|W_0)} \cdot 2^{-nH(Y_2|W_0)})$
$= 2^{-n(I(Y_2;W_1 W_2 U_2|W_0)+I(U_1;W_1|W_0)+I(W_1;W_2 U_2|W_0)-(T_1+t_2+z_2))}$

Similarly, the other terms can be evaluated. In order to $P_{e2}^n \to 0$ as the block length $n \to \infty$, it is necessary and sufficient that $(T_0, T_1, t_2, z_2)$ satisfy constrains in (A-1)-(A-15). As seen in (A-1)-(A-15), the inequalities A-1,5,6,7,11,12, 13 are redundant due to (A-15). Therefore, after removing these redundant inequalities and considering the binning conditions:
$t_2 - T_2 \geq I(W_2; W_1 U_1|W_0)$ or $T_2 - t_2 \leq -I(W_2; W_1 U_1|W_0)$
$z_2 - S_2 \geq I(U_2; U_1 W_1 W_2|W_0)$ or
$\qquad\qquad\qquad S_2 - z_2 \leq -I(U_2; U_1 W_1 W_2|W_0)$
we reach to the constraints (3-9)-(3-16). From a similar error analysis for the receiver $RX_1$, the inequalities (3-1)-(3-8) in theorem 1 are obtained.

## APPENDIX B
## PROOF OF THEOREM 2

We set $S_1 = R_1 - T_1$ and $S_2 = R_2 - T_2$ in all of the relations (3-1)-(3-16) in theorem 1. Then by using the Fourier-Motzkin elimination technique as in [15] and eliminating redundant relations, we reach to the constraints in theorem 2 (for brevity the details are omitted).

$$\begin{cases}
T_0 \leq I(Y_2; W_0 W_1 W_2 U_2) + I(U_2; W_2|W_0) + I(W_1; W_2 U_2|W_0) & \text{(A-1)}\\
T_1 \leq I(Y_2; W_1|W_0 W_2 U_2) + I(U_2; W_2|W_0) + I(W_1; W_2 U_2|W_0) & \text{(A-2)}\\
t_2 \leq I(Y_2; W_2|W_0 W_1 U_2) + I(U_2; W_2|W_0) + I(W_1; W_2 U_2|W_0) & \text{(A-3)}\\
z_2 \leq I(Y_2; U_2|W_0 W_1 W_2) + I(U_2; W_2|W_0) + I(W_1; W_2 U_2|W_0) & \text{(A-4)}\\
T_0 + T_1 \leq I(Y_2; W_0 W_1 W_2 U_2) + I(U_2; W_2|W_0) + I(W_1; W_2 U_2|W_0) & \text{(A-5)}\\
T_0 + t_2 \leq I(Y_2; W_0 W_1 W_2 U_2) + I(U_2; W_2|W_0) + I(W_1; W_2 U_2|W_0) & \text{(A-6)}\\
T_0 + z_2 \leq I(Y_2; W_0 W_1 W_2 U_2) + I(U_2; W_2|W_0) + I(W_1; W_2 U_2|W_0) & \text{(A-7)}\\
T_1 + t_2 \leq I(Y_2; W_1 W_2|W_0 U_2) + I(U_2; W_2|W_0) + I(W_1; W_2 U_2|W_0) & \text{(A-8)}\\
T_1 + z_2 \leq I(Y_2; W_1 U_2|W_0 W_2) + I(U_2; W_2|W_0) + I(W_1; W_2 U_2|W_0) & \text{(A-9)}\\
t_2 + z_2 \leq I(Y_2; W_2 U_2|W_0 W_1) + I(U_2; W_2|W_0) + I(W_1; W_2 U_2|W_0) & \text{(A-10)}\\
T_0 + T_1 + t_2 \leq I(Y_2; W_0 W_1 W_2 U_2) + I(U_2; W_2|W_0) + I(W_1; W_2 U_2|W_0) & \text{(A-11)}\\
T_0 + T_1 + z_2 \leq I(Y_2; W_0 W_1 W_2 U_2) + I(U_2; W_2|W_0) + I(W_1; W_2 U_2|W_0) & \text{(A-12)}\\
T_0 + t_2 + z_2 \leq I(Y_2; W_0 W_1 W_2 U_2) + I(U_2; W_2|W_0) + I(W_1; W_2 U_2|W_0) & \text{(A-13)}\\
T_1 + t_2 + z_2 \leq I(Y_2; W_1 W_2 U_2|W_0) + I(U_2; W_2|W_0) + I(W_1; W_2 U_2|W_0) & \text{(A-14)}\\
T_0 + T_1 + t_2 + z_2 \leq I(Y_2; W_0 W_1 W_2 U_2) + I(U_2; W_2|W_0) + I(W_1; W_2 U_2|W_0) & \text{(A-15)}
\end{cases}$$